# Enhancement of NMR Signals via Phase Whitening and Quantum Fourier Transform

**Toshio Fukumi**
Quantum Genome Informatics
1-2-30-207 Kiyoshikojin, Takarazuka, 665 0836, Japan
fukumit@zmail.plala.or.jp
**Shizuo Fujiwara**
Professor Emeritus in The University of Tokyo
1-8-2 Negishi, Saitama 336-0624, Japan
fujiwras@peach.ocn.ne.jp

The Zeeman energy to be considered in NMR spectroscopy is small compared to thermal energy, and it results in the poor sensitivity of NMR. We propose here a new method to enhance NMR signals by phase whitening followed by quantum Fourier transformation. This method will suggest that 14 spins NMR quantum computer can enhance NMR signal by the factor of about 10 and enhancement increases exponentially.

NMR is most powerful spectroscopy in chemistry and biochemistry. The main drawback of NMR spectroscopy is in its low sensitivity. The S/N ratio of the normal traditional CAT method could be enhanced by the factor of $O(\sqrt{N})$, where N refers to the number of interactions. As it will be shown below, the Fourier transformation can enhance S/N ratio efficient and significant manner.

It is known that the white noise can be defined by the characteristic functional

$$C(\xi) = \exp(-\frac{1}{2}\|\xi\|^2), \ \xi \in \varphi,$$

where $\varphi$ is a nuclear space, and let $\varphi^*$ be it's dual.[1] The elements of $\varphi^*$ are sample paths of white noise $\dot{B}(t)$. Denoting functional differential w.r.t. white noise as $\partial_t$ which is an annihilation operator and the dual $\partial_t^*$ is creation operator. Then we can denote white noise as

$$\dot{B}(t) = \partial_t + \partial_t^*$$

In physics language, we have

$$W(t) = a(t) + a^*(t)$$

Where $a$ and $a^*$ stand for annihilation and creation operators, respectively.

Apparently white noise can do no business as it annihilates and creates simultaneously, but it is not so here. Multiplication by white nose results in the phase change

$$W(t)\rho = e^{2\pi i\gamma}\rho,$$

where $\rho$ is a functional of Brownian motion, and $\gamma$ is a random number in $[0,1)$.

Before, we describe our strategy, le us note tat Fourier transform maps white noise into $\delta$ function,

$$F:W(t) \to \delta.$$

The key strategy is to phase whitening by the multiplication to spin and to employ QFT (Quantum Fourier Transformation).

Obviously, we cannot obtain NMR signals by spin whose phase is white. Classical circuits cannot make any business in it self because we can never pick up NMR signal from phase-whitened state. In this sense, the normal FFT technique like NOE (Nuclear Overhauser Effect) can be used in our case.

On the other hand QFT can ideally transform randomized quantum phase, into $\delta$ function. However this relays on the ideal white noise case. For the sake of materialization, we work, let us consider a practical case.

As one spin acts as qubit which means that we can manipulates 2 bits, a simple calculation indicates that 14 spins NMR quantum computer can enhances the conventional NMR's S/N ratio by the factor of about 10. Number of spins which will able to contribute NMR signals:

**Conventional NMR**
**Avogadro #→sample tube→Boltzman factor→solute**
$\quad\quad 10^{23} \rightarrow 10^{20} \rightarrow 10^{14} \rightarrow 10^{11}$

**Phase Whitening QFT NMR:**
$\quad\quad 2^{14} \rightarrow 10^{12}$ **spins.**

As the qubits increasers exponentially w.r.t the number of spins, which means S/N ration increase exponentially in the present scenario. It is tempting that Fourier transformation transforms frequency domain signals into time domain signals, and .vise versa.

But it should be noted that frequency signals are equivalent to phase modulated signals. The quantum Fourier transform was invented by Shor[2] in 1992 and was implemented into NMR quantum computer[3] and this algorithm play the main role in the present study in enhancing NMR signals

Quantum Fourier transform operation is defined by

$$U_{FT}|x> = \frac{1}{2}\sum_{y=0}^{2^n-1} e^{i(2\pi/2^n)xy}|y>.$$

This operation can be reduced to Hadammard transform,

$$|x_i> \xrightarrow{H} |0> + e^{i2\pi(x_i/2)}|1>,$$

and Controlled-Phase Shift gate,

$$|x_i>|x_j> \xrightarrow{R_m(i,j)} (|0> + e^{i2\pi(x_j/2^m)}|1>)|x_j>.$$

Our procedure is composed by target spin(s) to be enhanced and arithmetic spins which carry QFT.

Our scheme is summarized as
   **1. apply 90° pulse to the target spins.**
   **2. Gz-pulses for whitening of phase of target spins.**
   **3. entangle target spin with register.**
   **4. quantum Fourier transform.**
   **5. receiver coil and FFT.**

The crucial point is that the only target spins enjoy phase whitening as the spins are in transverse directions. It should be noted that FFT can not do the business as the receiver coil cannot pick up any change polarization of target spins via controlling the polarization of spin though dipole coupling and /or spin coupling. If apply the 90° pulse, the polarization of spins will increase dramatically. This can be used spin labeling for protein NMR spectroscopy without shortening transverse relaxation like electron spin-cross polarization.

    The most useful application of our method will be to employ NOE(Nuclear Overhauser Effect). The NOE can be used to determine chemical structures of molecules[4] and molecular mobility[5]. NOE is a most appreciated tool for determinationof structures of proteins.

    Finally, let us emphasize that present proposal will be the first practical application of a quantum information technology realizable experimentally.

    The detail will be appear in the forthcoming full paper.

References
1. T.Hida, "Brownian Motion" Springer, 1980.
2. P.W.Shor, SIAM J. Comput., **26**, 1484, (1997).
3. U.Sakaguchi. H.Ozawa, and T.Fukumi, "NMR Quantum Computing" in "Coherence and Statistics of Photons and Atoms" ed. J.Perina, J.Wiley (2001) and refences therein.
4. T.Fukumi, Y.Arata,and S.Fujiwara, J. Mol. Spectrosc. **27**, 443, (1968).
5. T.Fukumi, Y.Arata, and S.Fujiwara, J. Chem. Phys. **49**, 4198 (1968)